\documentclass[preprint]{ptephy_v1}

\preprintnumber{XXXX-XXXX} 
\usepackage{hyperref}

\begin{document}

\title{Delocalization and re-entrant localization of flat-band states in non-Hermitian disordered lattice models with flat bands}


\author[1]{Sangbum Kim}
\affil{Research Institute for Basic Sciences, Ajou University,
Suwon 16499, Korea \email{sangbumkim@ajou.ac.kr}}

\author[2,3]{Kihong Kim}
\affil{Department of
Physics, Ajou University, Suwon 16499, Korea \email{khkim@ajou.ac.kr}}
\affil[3]{School of Physics, Korea Institute for Advanced Study, Seoul 02455, Korea }


\begin{abstract}%
We present a numerical study of Anderson localization in disordered non-Hermitian lattice models with flat bands.
Specifically we consider one-dimensional stub and two-dimensional
kagome lattices that have a random scalar potential and a uniform imaginary vector potential and calculate the spectra of
the complex energy, the participation ratio, and the winding number as a function of the strength of the imaginary vector potential, $h$. The flat-band states are found to show a double transition
from localized to delocalized and back to localized states with $h$, in contrast to the dispersive-band states going through a single delocalization transition.
When $h$ is sufficiently small, all flat-band states are localized. As $h$ increases above a certain critical value $h_1$,
some pair of flat-band states become delocalized. The participation ratio associated with them increases substantially and their winding numbers
become nonzero. As $h$ increases
further, more and more flat-band states get delocalized until the fraction of the delocalized states reaches a maximum. For larger $h$ values,
a re-entrant localization takes place and, at another critical value $h_2$, all flat-band states return to
compact localized states with very small participation ratios and zero winding numbers.
This re-entrant localization transition, which is due to the interplay among disorder, non-Hermiticity, and flat band, is a
phenomenon occurring in many models having an imaginary vector potential and a flat band simultaneously. We explore the spatial characteristics of
the flat-band states by calculating the local density distribution.
\end{abstract}

\subjectindex{I50 Disordered systems, Anderson transitions, I9 Low dimensional systems - electronic properties}

\maketitle

\section{Introduction}

Even though Anderson localization of quantum particles and classical waves in random media has been studied extensively for over six decades \cite{pwa,lee,eve,gre,seg}, many aspects of it are not yet fully understood and localization phenomena with qualitatively new characteristics are still being discovered when
adding new ingredients to the model \cite{yus,segev,iomin,kim4,kim5,suz,long3}. In this paper, we explore a novel type of Anderson localization that occurs
when the system is non-Hermitian and has a dispersionless flat band in its energy spectrum at the same time.

Non-Hermiticity can be introduced into the model in various ways such as adding an imaginary scalar or vector potential \cite{nhq,ash}.
The localization-delocalization transition
in one- and two-dimensional lattice models with a random real scalar potential and a constant imaginary vector potential
has been studied extensively since first being proposed by Hatano and Nelson \cite{hn,hat3,shn,kim1,ref,long1,long2}. Here we generalize such studies to the
lattices supporting a flat band. There has been substantial recent interest in the properties
of electronic and photonic systems with flat bands \cite{ley,luck,lim,bal}. In those systems, the group velocity associated with the flat band vanishes which
causes many interesting phenomena to occur by enhancing
the effects of various perturbations including disorder \cite{goda,chalk,ley2,ley3,shu1,kim2}.
Two-dimensional (2D) lattices such as the Lieb, dice, and kagome lattices and one-dimensional (1D) lattices
such as the stub, sawtooth, and diamond lattices are among the representative examples exhibiting flat bands \cite{zong,wei,real,liu}.
Our main aim in this paper is to investigate the influence of non-Hermiticity due to the imaginary vector potential $\bf g$ on the localization
of the flat-band states. Especially we will demonstrate that as the strength of $\bf g$ increases,
a substantial portion of those states go through a double transition from localized to delocalized and then back to localized states.
This type of re-entrant localization transition has never been studied previously.

\section{Model and method}

\begin{figure}
\centering\includegraphics[width=6cm]{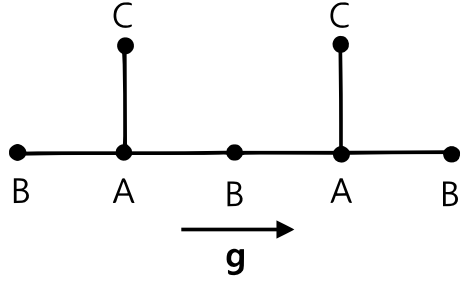}
\caption{\label{fig:epsart1} Sketch of the stub lattice. The direction of the imaginary vector potential $\bf g$ is indicated with an arrow.}
\label{fig:sl}
\end{figure}

We generalize the Hatano-Nelson model
to the cases with a flat band such as the stub lattice, a schematic of which is shown in Fig.~\ref{fig:sl}.
We notice that a unit cell of the stub lattice consists of three inequivalent sites, A, B, and C.
The stationary discrete Schr\"{o}dinger equation for the stub lattice can be written as
\begin{eqnarray}
E_\lambda \psi_{\lambda}({\rm A},n)&=& v_n^{\rm A}\psi_{\lambda}({\rm A},n)+te^h\psi_{\lambda}({\rm B},n-1)\nonumber\\&&+te^{-h}\psi_{\lambda}({\rm B},n)+t\psi_{\lambda}({\rm C},n),    \nonumber\\
E_\lambda \psi_{\lambda}({\rm B},n)&=& v_n^{\rm B}\psi_{\lambda}({\rm B},n)+te^h\psi_{\lambda}({\rm A},n)\nonumber\\&&+te^{-h}\psi_{\lambda}({\rm A},n+1),    \nonumber\\
E_\lambda \psi_{\lambda}({\rm C},n)&=& v_n^{\rm C}\psi_{\lambda}({\rm C},n)+t\psi_{\lambda}({\rm A},n),
\label{eq:main}
\end{eqnarray}
where $\psi_{\lambda}(m,n)$ ($m={\rm A}$, B, C) represents the eigenfunction at the $n$-th unit cell and $E_\lambda$ is the (complex) energy eigenvalue.
$v_n^m$ is an on-site random potential chosen from the uniform distribution in $[ -\Delta,\Delta]$.
If we assume $\bf g$ to have the dimension of momentum, then
the dimensionless non-Hermiticity parameter $h$ is defined by $h=ga/\hbar$, where $a$ is the lattice spacing. $te^h$ and $te^{-h}$ represent the asymmetric nearest-neighbor hopping
integrals between the A and B sites.

We apply the periodic boundary condition to a stub lattice with total unit cell number $N_c$ and total site number $N$ ($=3N_c$) and
solve the eigenvalue problem given by Eq.~(\ref{eq:main}). Since the system is non-Hermitian, we need to calculate both the {\it normalized} left and right eigenfunctions
$\psi_\lambda^L(m,n)$ and $\psi_\lambda^R(m,n)$ corresponding to the same eigenvalue $E_\lambda$, using
which we compute the participation ratio $P_\lambda$ defined by
\begin{eqnarray}
P_\lambda = \frac{1} {\Sigma_{m}\Sigma_{n=1}^{N_c} \big\vert \psi_\lambda^L(m,n) \psi_\lambda^R(m,n) \big\vert^2}.
\end{eqnarray}
We note that $P_\lambda$ is bounded by $1\le P_\lambda\le N$.

Next we calculate the winding number $W_\lambda$, which is defined in the same way as in \cite{shn,kim1}:
\begin{eqnarray}
W_\lambda=\frac{1}{2\pi}\oint_{C_0} d\alpha_\lambda,
\end{eqnarray}
where $\alpha_\lambda$ is the site-dependent phase of the eigenfunction and $C_0$ is a closed contour
surrounding the origin of the complex plane.
When the eigenvalue $E_\lambda$ is complex, the eigenfunctions $\psi_\lambda^R(m,n)$ and
$\psi_\lambda^L(m,n)$ rotate (in the opposite directions) an integer times around the origin on the complex plane,
as the closed contour is traversed once.
In actual calculations, we choose any closed contour around a ring of the stub lattice (satisfying the periodic boundary condition) 
and sum the phase of the eigenfunction $\psi_\lambda^R(m,n)$
[or equivalently $\psi_\lambda^L(m,n)$] at every site on the contour and divide the result by $2\pi$,
when the phase at a reference site is fixed to zero. Among many possible contours, we choose the simplest one traversing only the A and B sites. The present winding number, which is well-defined even in strongly
disordered cases, is the generalization for non-Hermitian disordered
systems of the wave vectors labelling the eigenstates in non-disordered systems. This quantity needs to be
distinguished from the winding number describing the vorticity of complex energy eigenvalues
\cite{gong,yoshida}.

The spectrum of a clean stub lattice (i.e., $\Delta=0$) with an imaginary vector potential $h$
is given by
\begin{eqnarray}
E_\lambda=\left\{\begin{array}{l l}
\pm \sqrt{3+2\cosh\left(2h-iq_\lambda\right)}\\
0
\end{array}\right.,
\end{eqnarray}
where $q_\lambda$ ($=2\pi\lambda/N_c$, $\lambda=1, 2, \cdots, N_c$) is a wave vector in the first Brillouin zone \cite{luck}.
It consists of a highly degenerate flat band with $N_c$ states at $E_\lambda =0$ and
either two separate dispersive bands each with $N_c$ states (for $h\lesssim 0.48$)
or one merged dispersive band with $2N_c$ states (for $h \gtrsim 0.48$).
The (unnormalized) flat-band eigenfunction is a compact localized state in which it is nonzero only at three adjacent sites
\begin{eqnarray}
&&\psi_{\lambda}({\rm C},n)=e^{-h},~~\psi_{\lambda}({\rm C},n+1)=e^{h},\nonumber\\
&&\psi_{\lambda}({\rm B},n)=-1.
\label{eq:wf}
\end{eqnarray}
We note that $P_\lambda=3$ when $h=0$ and $P_\lambda\approx 1$ when $h$ is large.
When weak disorder is turned on, the clear distinction between flat and dispersive bands is maintained
except in the very narrow transition region near and slightly below $h= 0.48$.
In such cases, we define $P_{\rm FB}$ as the participation ratio averaged over $N_c$ flat-band states and $P_{\rm av}$ as
that averaged over all $N$ states. We also consider many independent disorder configurations with the same $\Delta$ and denote the ensemble average as $\langle\cdots\rangle$.

\begin{figure}
\centering\includegraphics[width=\textwidth]{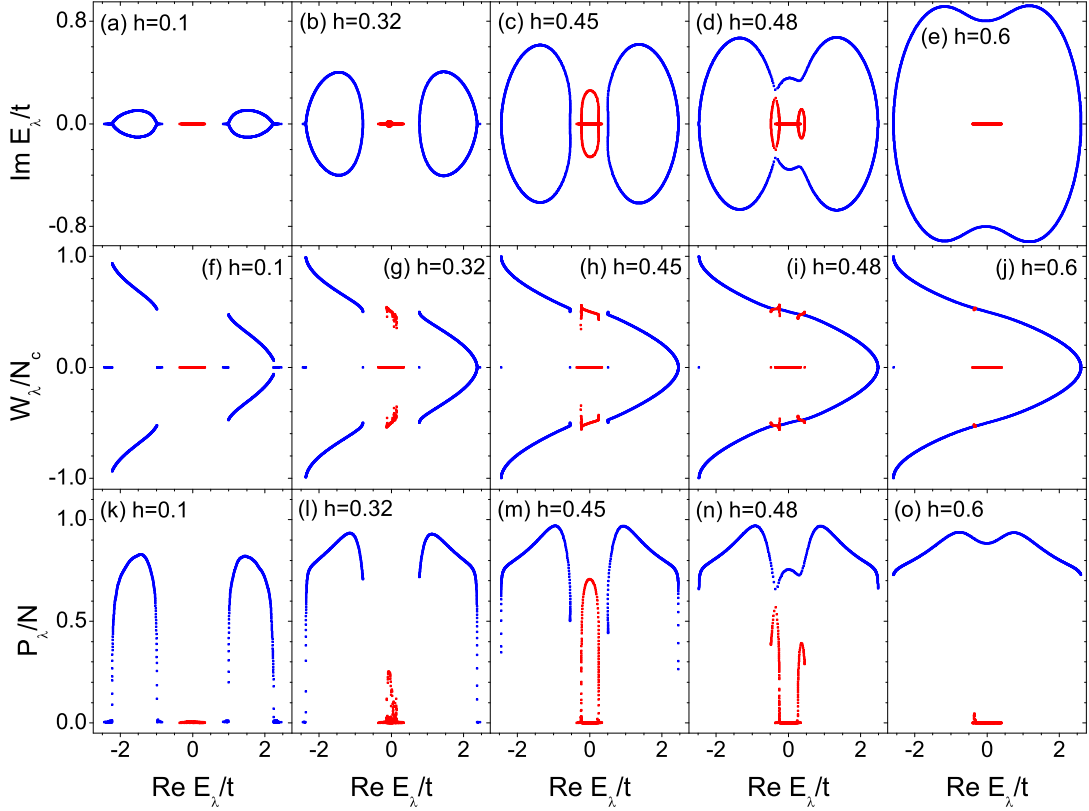}
\caption{(a-e) Complex energy spectra of a stub lattice in a single disorder configuration when $N_c=800$, $N(=3N_c)=2400$, $\Delta/t=0.4$, and
$h=0.1$, 0.32, 0.45, 0.48, 0.6. (f-j) Winding number $W_\lambda$ divided by the total unit cell number $N_c$ and (k-o) the participation ratio $P_\lambda$
divided by the total site number $N$ plotted versus the real part of the energy eigenvalue normalized by $t$. The blue (red) color corresponds to the dispersive bands (flat band) consisting of
1600 (800) states.}
\label{fig:spec}
\end{figure}

\section{Numerical results}

\subsection{Stub lattice}

In Fig.~\ref{fig:spec}, we show our numerical results for the spectra of the energy eigenvalue, the winding number, and the participation ratio
obtained for a stub lattice in a single disorder configuration, when $N_c=800$, $\Delta/t=0.4$, and
$h=0.1$, 0.32, 0.45, 0.48, 0.6. We plot the results for the flat band in red and those for the dispersive bands in blue.
The flat-band states show markedly different behavior from the dispersive-band states.
The evolution of the spectra for the dispersive bands with $h$ when $h\lesssim 0.47$ is qualitatively similar to that for the 1D Hatano-Nelson model.
When $h$ is zero or sufficiently small, all states are localized with real energy eigenvalues and zero winding numbers.
As $h$ increases above a small critical value $h_{a}$, which is about 0.007 in the present disorder configuration,
some states near the centers of the two dispersive bands get delocalized and their eigenvalues become complex.
Delocalization always occurs for a pair of states
with the eigenvalues being the complex conjugate of each other. When there are enough delocalized states, the plot of their eigenvalues in the complex plane takes the shape
of a {\it bubble}. As $h$ increases further above another critical value $h_b$ ($\sim 0.45$),
all states in the dispersive bands become delocalized and their eigenvalues become complex except for those at the band edges.
In Fig.~\ref{fig:spec}(c), we note that two large (blue) bubbles corresponding to the two dispersive bands are formed.

When delocalization occurs for a pair of states, the winding number changes from zero to a pair of nonzero values
with the same magnitude but of opposite sign.
We observe that for dispersive-band states with nonzero $W_\lambda$, there is no degeneracy of $W_\lambda$ and its absolute
value is a monotonically decreasing function of ${\rm Re}~E_\lambda$.
The values of $\vert W_\lambda\vert$ for the lower dispersive band are larger than
those for the higher one.
When all dispersive-band states are delocalized,
the range of $W_\lambda$ is $0\le \vert W_\lambda\vert \le N_c-1$.
The winding numbers for all band-edge states remain zero.
The participation ratios for most delocalized states are much larger than those for localized ones. The maximum of
$P_\lambda$ is approximately $0.97N$. This means that the most homogeneous delocalized state in the dispersive bands is the one spreading almost evenly over all lattice sites.

\begin{figure}
\centering\includegraphics[width=12cm]{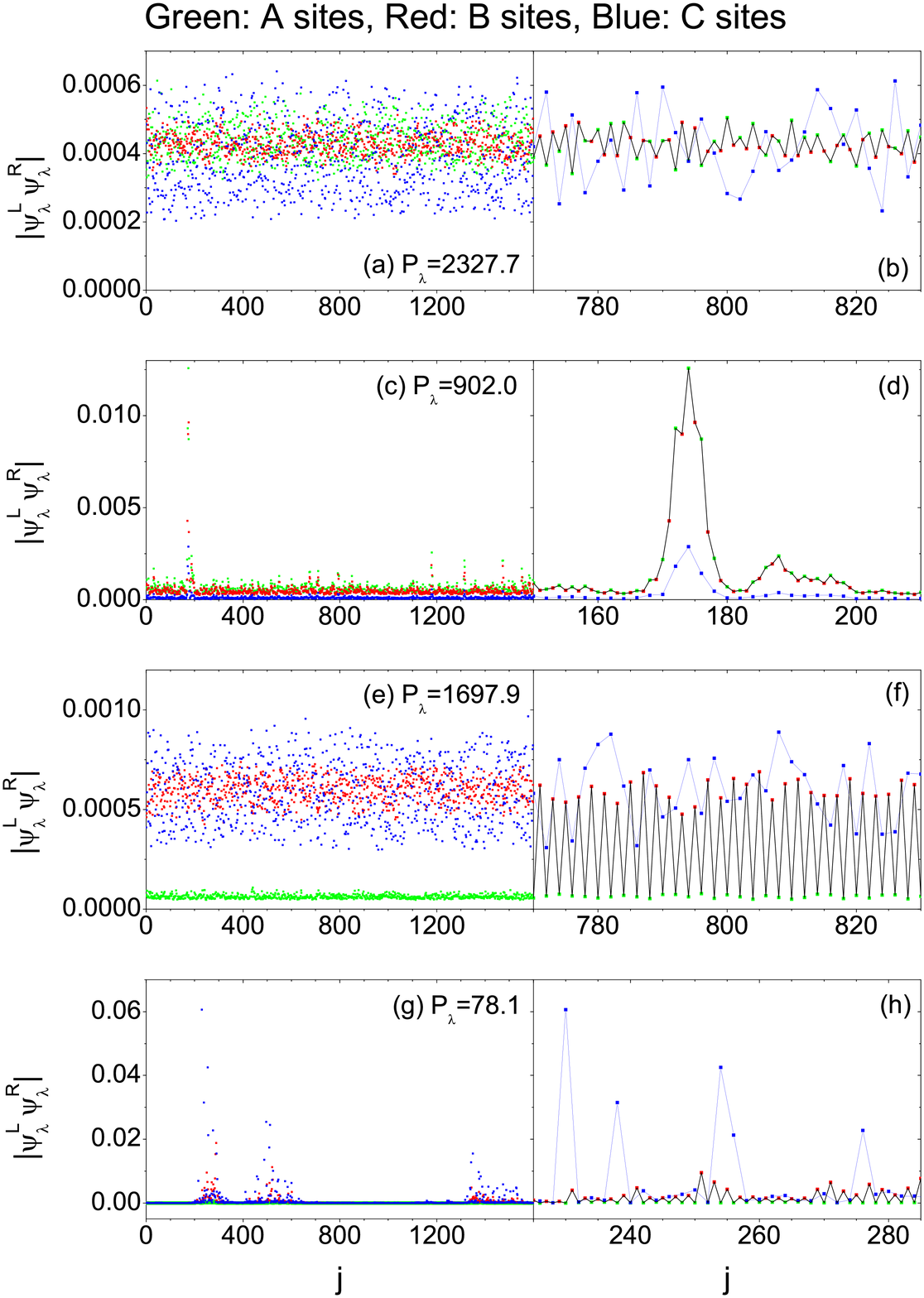}
\caption{Local density distributions for (a-d) two typical dispersive-band and (e-h) two typical flat-band states plotted versus $j$,
the odd [even] values of which correspond to the {\rm B} (red) [{\rm A} (green) and {\rm C} (blue)] sites in the same disorder configuration as in Fig.~\ref{fig:spec},
when $N_c=800$, $\Delta/t=0.4$, and $h=0.45$. (b), (d), (f), and (h) are the magnified versions of (a), (c), (e), and (g) respectively.
In (a), (c), (e), and (g), $P_\lambda$ is equal to 2327.7, 902.0, 1697.9, and 78.1 respectively.}
\label{fig:wf}
\end{figure}

The evolution of the spectra for the flat band can be seen from the red-colored parts of Fig.~\ref{fig:spec}.
When $h$ is smaller than $h_1$, which is about 0.26 in the present configuration and much larger than $h_a$ corresponding to the dispersive bands, all flat-band states are localized with real eigenvalues, zero winding numbers, and
very small participation ratios of $P_\lambda\gtrsim 3$.
As $h$ increases above $h_1$,
some states near the center of the flat band get delocalized and their eigenvalues become complex, similarly to the dispersive-band case.
Again, delocalization occurs for a pair of states
with the eigenvalues being the complex conjugate of each other. As $h$ increases further,
the plot of the delocalized eigenvalues in the complex plane is characterized by a (red) bubble as in Fig.~\ref{fig:spec}(c).
However, unlike in the dispersive case where all states within the energy range of a bubble are delocalized,
a majority of the states inside the range of a bubble remain localized in the flat-band case. Therefore the spectrum takes the shape
of a {\it bubble
pierced by the real line}. The fraction of delocalized states among all flat-band states is about 0.29 in Fig.~\ref{fig:spec}(c).
Most of the winding numbers for the delocalized states take the values close to $N_c/2$. We observe that for each of almost all the flat-band states
with nonzero $W_\lambda$, there exists a dispersive-band state with the same winding number. Furthermore, there are a few cases where multiple
flat-band states have the same winding number.
The participation ratios of delocalized flat-band states can be substantially large.
We find that the maximum
$P_\lambda$ for the flat band is about $0.71N$, which implies that the most homogeneous delocalized state is the one spreading almost evenly over the two-thirds of the sites. We will show later in Fig.~\ref{fig:wf} that only the B and C sites are occupied in this case.

\begin{figure}[h]
\centering\includegraphics[width=12cm]{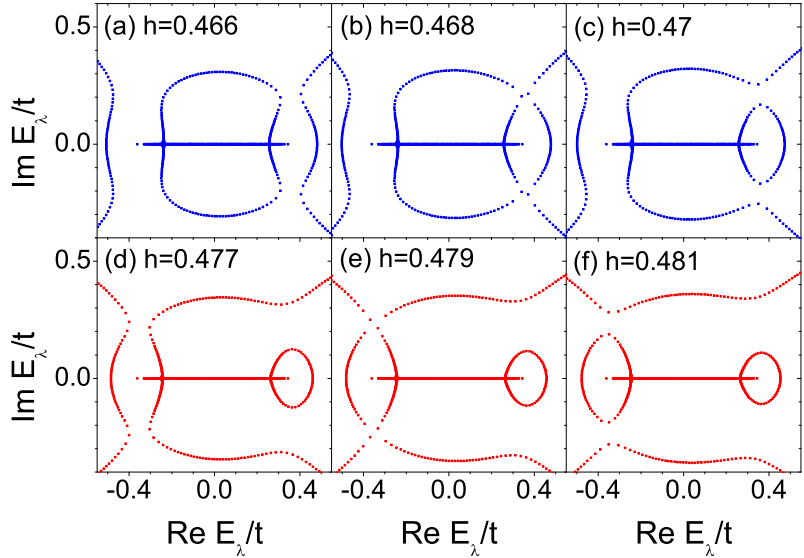}
\caption{Evolution of the complex energy spectrum of a stub lattice as $h$ increases (a-c) from 0.466 to 0.47 and (d-f) from 0.477 to 0.481 in the same disorder configuration as in Fig.~2
when $N=2400$ and $\Delta/t=0.4$.}
\label{fig:supp1}
\end{figure}

In the disorder configuration considered in Fig.~\ref{fig:spec},
the flat- and dispersive-band states go through merging and decoupling in two steps
within a very narrow range of $h$ between 0.467 and 0.48, as shown in Fig.~\ref{fig:supp1}.
More specifically, at $h\approx 0.468$, the flat-band bubble
merges with the right dispersive-band bubble and then gets separated as a smaller bubble in the positive energy region when $h\approx 0.469$. At $h\approx 0.479$, another merging with the left dispersive-band bubble occurs and then the decoupling of the flat and dispersive bands is completed when $h\approx 0.48$, at which
the flat-band spectrum consists of two bubbles connected by a real line as shown in Fig.~\ref{fig:supp1}(f).
Above $h=0.48$, the dispersive-band bubbles
are merged into one big bubble and are separated from the flat band.
The width of the region of $h$ where the merging and decoupling occurs increases gradually as the
the disorder strength $\Delta$ increases.
As $h$ increases above 0.48, the effect of the imaginary vector potential
begins to dominate the disorder effect and
more and more states become re-localized with real eigenvalues, zero winding numbers, and very small participation ratios. At $h\approx 0.63$, all flat-band states return to compact localized states with $P_\lambda\approx 1$.

\begin{figure}
\centering\includegraphics[width=10cm]{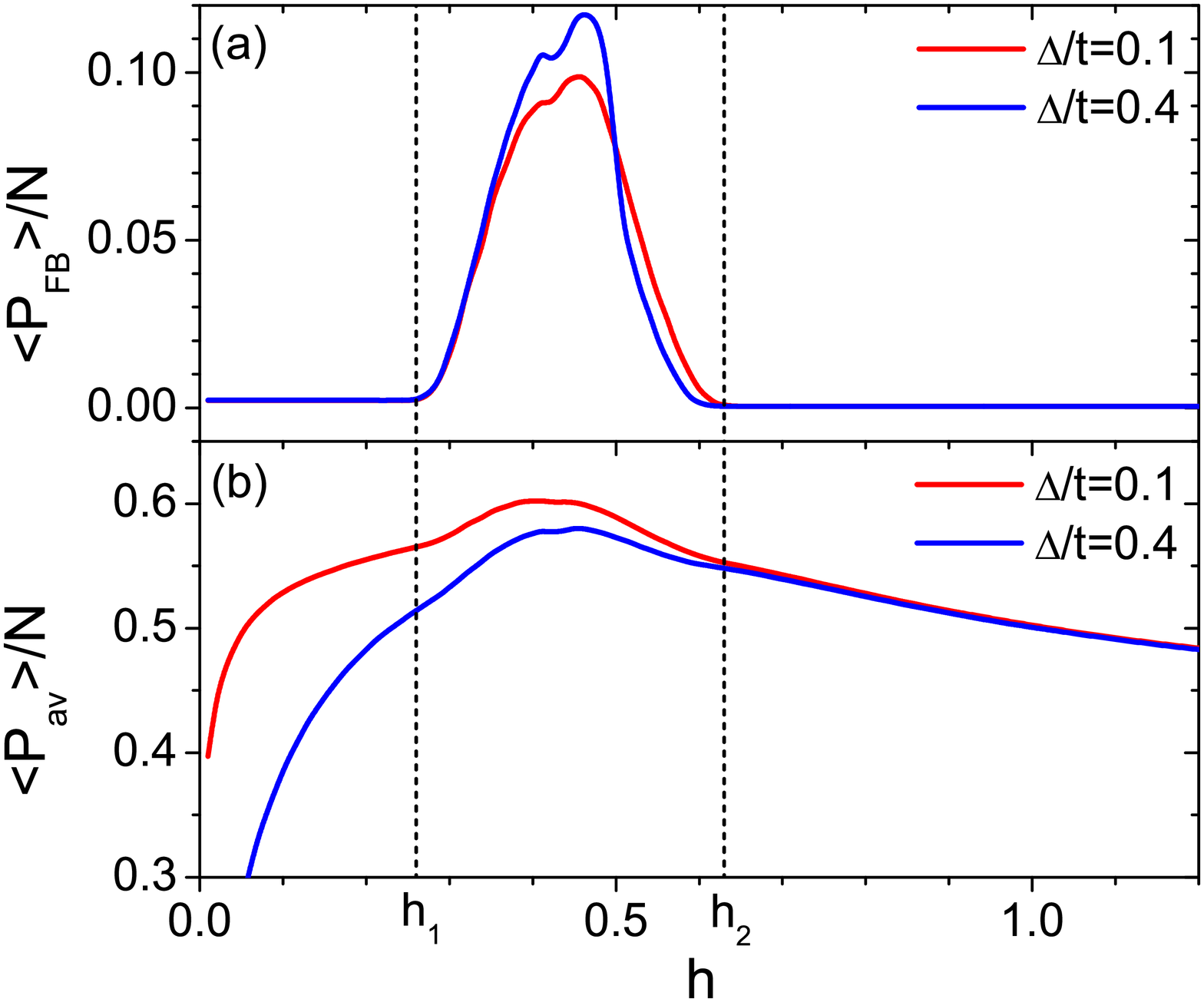}
\caption{Normalized participation ratios of a stub lattice averaged over (a) 800 flat-band states and (b) all 2400
states and averaged over 100 different disorder configurations
when $N=2400$ and $\Delta/t=0.1$, 0.4 plotted versus $h$. The positions of $h_1$ and $h_2$, between which the delocalization of some flat-band states occurs,
are indicated by dashed lines.}
\label{fig:pr}
\end{figure}

In order to show the double transition from localized to delocalized and back to localized flat-band states more explicitly,
we plot $\langle P_{\rm FB}\rangle/N$ and $\langle P_{\rm av}\rangle/N$ obtained by averaging over 100 independent disorder configurations versus $h$,
when $N=2400$ and $\Delta/t=0.1$ and 0.4, in Fig.~\ref{fig:pr}.
The behavior of $\langle P_{\rm FB}\rangle$ shows unambiguously the delocalization transition at $h=h_1$ and the re-entrant
localization transition at $h=h_2$.
For $h<h_1$, $\langle P_{\rm FB}\rangle$ remains very close to 3
corresponding to a compact localized state at three adjacent B and C sites.
For $h>h_2$, $\langle P_{\rm FB}\rangle$ is close to 1 corresponding to a compact localized state at a single
C site. Between $h_1$ and $h_2$, $\langle P_{\rm FB}\rangle$ increases rapidly, takes a maximum value
near $h\approx 0.45$, and then decreases sharply to 1. The maximum of $\langle P_{\rm FB}\rangle/N$
depends on the disorder and is about 0.12 when $\Delta/t=0.4$.
From Fig.~\ref{fig:pr}(b), we find that the double transition of the flat-band states noticeably affects the shape of
the $\langle P_{\rm av}\rangle$ curve, which is characterized by the appearance of
of a distinct peak in the region between $h_1$ and $h_2$,
though its $h$ dependence is not as dramatic as $\langle P_{\rm FB}\rangle$.

In Fig.~\ref{fig:wf}, we plot local density distributions for
two typical dispersive-band and two typical flat-band states versus $j$,
the odd [even] values of which correspond to the {\rm B} (red) [{\rm A} (green) and {\rm C} (blue)] sites in the same disorder configuration as in Fig.~\ref{fig:spec},
when $N_c=800$, $\Delta/t=0.4$, and $h=0.45$. Figs.~\ref{fig:wf}(b), (d), (f), and (h) are
the magnified versions of (a), (c), (e), and (g) respectively.
In the dispersive-band case, we find that the local density varies smoothly
between A and B sites, while the density at the C sites either fluctuates more strongly than or shows a different spatial dependence
from those at the A and B sites. In the flat-band case,
the local density at the A sites always remains extremely small, while those at the B and C sites are sizable.
For the flat-band states with large participation ratios such that $P_\lambda/N\approx 0.7$, all B and C sites
are roughly equally populated. For those with small participation ratios,
only a small number of C sites are substantially occupied as shown in Fig.~\ref{fig:wf}(h).
These behaviors can be understood from the fact
that the wave functions for the flat-band states are obtained approximately by a superposition of the functions
given by Eq.~(\ref{eq:wf}) for many different $n$'s, all of which have zero amplitudes at the A sites.

\subsection{Kagome lattice}

We have also considered other lattices with flat bands, such as the kagome lattice in 2D and the sawtooth lattice in 1D.
In all cases, we have obtained qualitatively similar results
to those for the stub lattice. In this subsection, we present the results for the kagome lattice.

\begin{figure}
\centering\includegraphics[width=6cm]{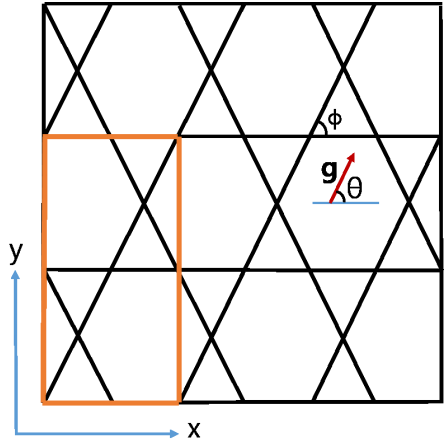}
\caption{Sketch of the kagome lattice. The direction of the imaginary vector potential $\bf g$ is indicated with an arrow.
The unit cell considered in the numerical calculation is shown in the orange box. }
\label{fig:kago}
\end{figure}

In Fig.~\ref{fig:kago}, we show a sketch of the kagome lattice considered in this paper.
The kagome lattice is a triangular lattice with a basis of three inequivalent sites where
each lattice site has four nearest neighbors. It is known that the band structure of this lattice consists
of two dispersive bands and one flat band. The shape of the lattice depends on the angle $\phi$.
In the present work, we fix $\phi=60^\circ$ and then all the nearest neighbor distances are the same and equal to $a$.
For the convenience of numerical calculations, we consider the (conventional) unit cell shown in the orange box.
This cell contains
six points and six different directions for the nearest-neighbor hopping.
We repeat these unit cells $N_x$ and $N_y$ times in the $x$ and $y$ directions respectively. Then the total number of sites $N$
is equal to $6N_xN_y$. The effective Hamiltonian describing the kagome lattice can be written as
\begin{eqnarray}
\mathcal{H}=-\frac{t}{2}\sum_{\langle i,j\rangle}\left(e^{{\bf g}\cdot{\bf s}_{ij}/\hbar} b^\dagger_{i}b_{j}
+e^{-{\bf g}\cdot{\bf s}_{ij}/\hbar} b^\dagger_{j}b_{i}\right)+\sum_{i}v_{i}b^\dagger_{i}b_{i},
\label{eq:ham}
\end{eqnarray}
where $b_i^\dagger$ and $b_i$ are the creation and annihilation operators at the site $i$ respectively
and ${\bf s}_{ij}$ denotes the lattice vector directed from the site $j$ to the site $i$. The notation $\langle i,j\rangle$
denotes that the summation is taken over only the nearest-neighbor sites.
$\bf g$ is an imaginary vector potential and $v_i$ is an on-site random potential chosen from the uniform distribution in $[-\Delta,\Delta]$.
We fix the direction of $\bf g$ so that $\theta=45^\circ$.
Then the quantity ${\bf g}\cdot{\bf s}_{ij}/\hbar$ takes six different values $\pm h/\sqrt{2}$, $\pm (\sqrt{6}+\sqrt{2})h/4$, and $\pm (\sqrt{6}-\sqrt{2})h/4$ where $h=ga/\hbar$, depending on
the choice of ${\bf s}_{ij}$.

\begin{figure}
\centering\includegraphics[width=\textwidth]{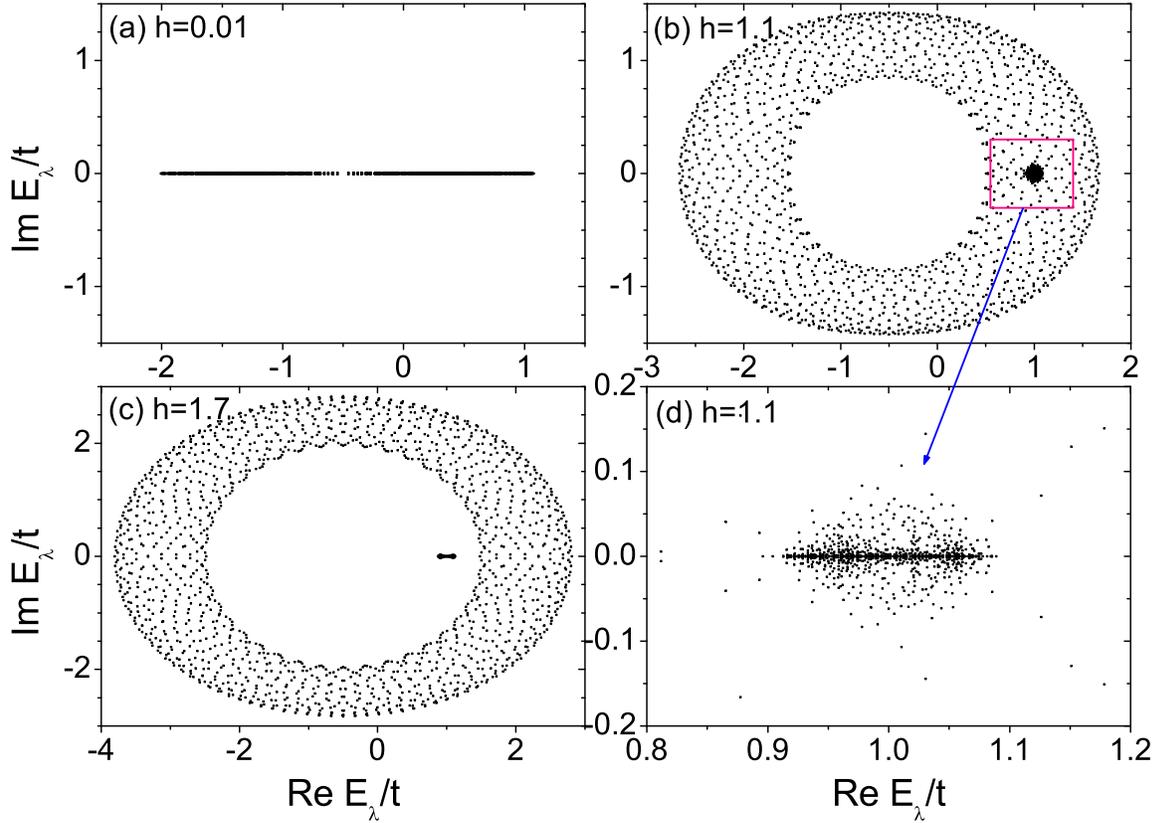}
\caption{(a-c) Complex energy spectra of a kagome lattice in a single disorder configuration when $N_x=N_y=20$, $N=6N_xN_y=2400$, $\Delta/t=0.1$, and
$h=0.01$, 1.1, 1.7. (d) An expanded view of the boxed region in (b).}
\label{fig:ks}
\end{figure}

In Fig.~\ref{fig:ks}, we show the complex energy spectra of a kagome lattice in a single disorder configuration,
when $N_x=N_y=20$, $N=6N_xN_y=2400$,  $\Delta/t=0.1$, and
$h=0.01$, 1.1, 1.7.
When $h$ is as small as $0.01$, all eigenstates are localized and their eigenvalues are real.
In the absence of disorder and the imaginary vector potential, the flat band of
the kagome lattice described by Eq.~(\ref{eq:ham}) is located precisely at the energy $E=t$. As the disorder is introduced, this band gets broadened. When $\Delta/t$ is equal to 0.1, the flat-band
states exist roughly in the range of $0.9<E/t<1.1$.

As $h$ increases, more and more states become delocalized and take complex energy eigenvalues.
At $h=1.1$, we find that all dispersive-band states are delocalized. Except for a small number of the band-edge states
with real eigenvalues, these states have
complex eigenvalues as shown in Fig.~\ref{fig:ks}(b). In contrast, however, the flat-band states are
only partially delocalized and a substantial portion of them remain localized for all $h$. An
expanded plot of the distribution of the eigenvalues inside the boxed region in Fig.~\ref{fig:ks}(b)
is shown in Fig.~\ref{fig:ks}(d). We clearly find that some fraction of the flat-band states
are delocalized and the distribution of their eigenvalues is markedly different from that in the dispersive-band case.
That most of the states with complex eigenvalues lying close to the real line in the range $0.9<{\rm Re}~E_\lambda/t<1.1$ belong to
the flat band is evident from the observation that there is no similar behavior near $-1.1<{\rm Re}~E_\lambda/t<-0.9$.
As $h$ increases further, the delocalized flat-band states begin to return to compact localized states, similarly to the behavior
of the stub lattice which
we have described in detail in the main text. When $h=1.7$, we find that almost all flat-band states
are re-localized with real energy eigenvalues.

\begin{figure}
\centering\includegraphics[width=8cm]{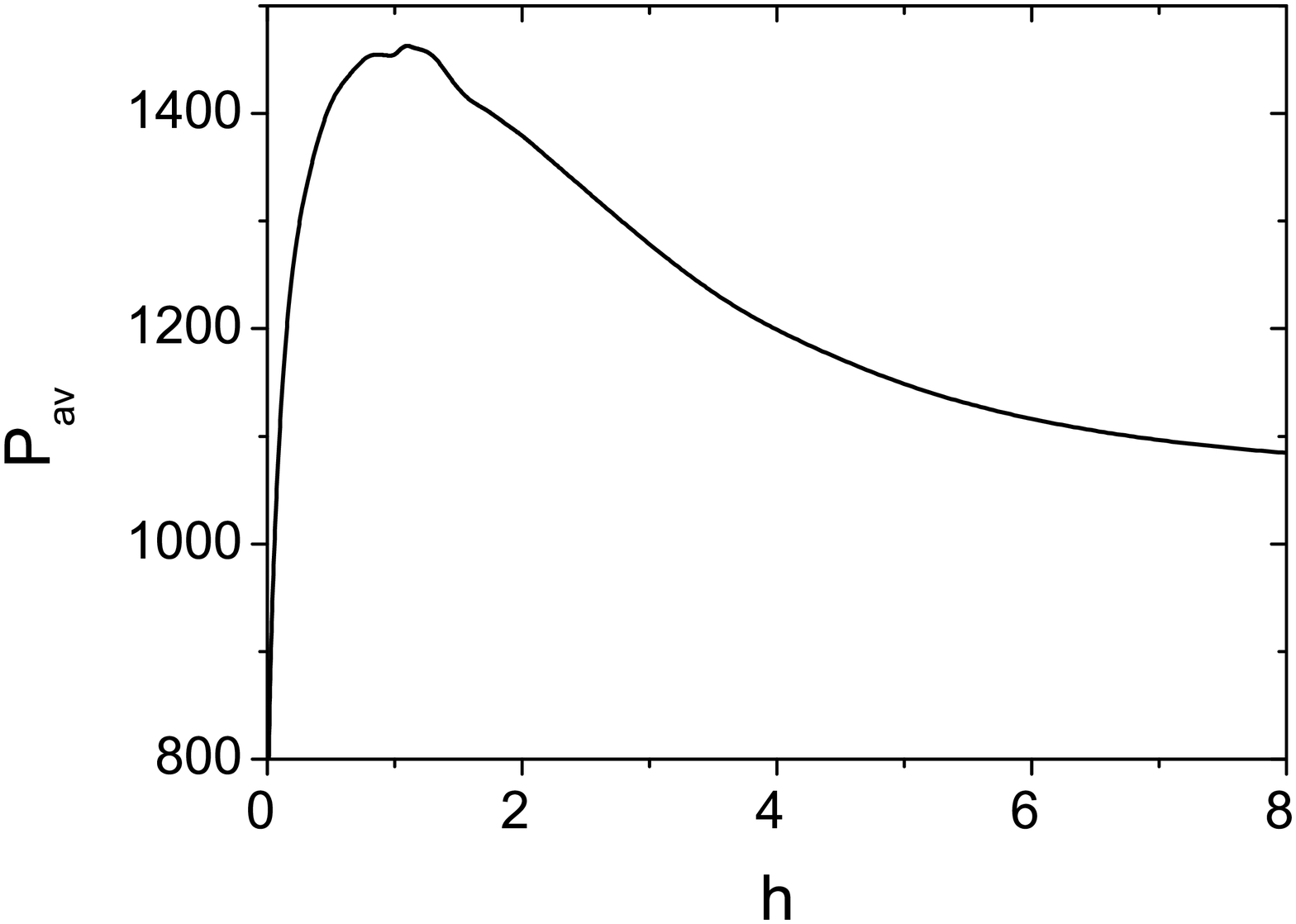}
\caption{Participation ratio of a kagome lattice averaged over all 2400 eigenstates in the same disorder configuration as in Fig.~\ref{fig:ks} when $N_x=N_y=20$ and $\Delta/t=0.1$ plotted versus $h$.}
\label{fig:kpr}
\end{figure}

In the kagome lattice, it is difficult to distinguish the flat and dispersive bands unambiguously
and therefore we have not calculated $P_{\rm FB}$.
However, we can observe the influence of the double transition
occurring to some flat-band states from the shape of the $P_{\rm av}$ curve.
In Fig.~\ref{fig:kpr}, we plot the participation ratio of a kagome lattice averaged over all eigenstates in the same disorder configuration as in Fig.~\ref{fig:ks} when $N_x=N_y=20$ and $\Delta/t=0.1$ versus $h$.
We find that this quantity takes a maximum value at $h\approx 1.1$.
The appearance of a clear peak of $P_{\rm av}$ as a function of $h$ can be considered as a consequence of the double transition
and is fully consistent with the behavior obtained for the stub lattice.

\section{Conclusion}

In conclusion, we have performed a numerical study of Anderson localization
in disordered non-Hermitian lattice models with flat bands.
We have found that there appears a double transition from localized to delocalized and back to localized states
for a substantial portion of the flat-band states and explored its characteristics in detail.
The re-entrant localization transition, which is due to the interplay among disorder, non-Hermiticity, and flat band, is a
phenomenon occurring in many disordered models with an imaginary vector potential and a flat band.

The predictions made in this study can be readily verified through experiments performed on a variety of systems
including ultracold atoms in optical lattices \cite{gou}, non-Hermitian electric circuit lattices \cite{hel}, superconducting vortex lattices \cite{bis}, and
synthetic photonic lattices constructed using optical waveguide arrays \cite{pob} or arrays
of coupled ring resonators \cite{liwang}, where the structures are fabricated to have flat bands.
The imaginary vector potential, or equivalently, the nonreciprocal hopping can be realized using methods such as Aharonov-Bohm rings \cite{gou,shap}
and coupled-resonator optical waveguide structures \cite{long1}.
Future experimental work in that direction will be of great interest.

\section*{Acknowledgment}

This research was supported through a National Research
Foundation of Korea Grant (NRF-2022R1F1A1074463)
funded by the Korean Government.
It was also supported by the Basic Science Research Program funded by the Ministry of Education (2021R1A6A1A10044950).

%

\vspace{0.2cm}
\noindent

\let\doi\relax


\end{document}